# Quantum teleportation between light and matter

Jacob F. Sherson[1,3], Hanna Krauter[1], Rasmus K. Olsson[1], Brian Julsgaard[1], Klemens Hammerer[2], Ignacio Cirac[2] & Eugene S. Polzik[1]

[1]Niels Bohr Institute, Copenhagen University, Blegdamsvej 17, Copenhagen Ø, Denmark. [2]Max Planck Institute for Quantum Optics, Hans-Kopfermann-Str. 1, Garching, D-85748, Germany. [3]Department of Physics and Astronomy, University of Aarhus, Aarhus, 8000, Denmark.

**Quantum teleportation[1] is an important ingredient in distributed quantum networks[2], and can also serve as an elementary operation in quantum computers[3]. Teleportation was first demonstrated as a transfer of a quantum state of light onto another light beam[4–6]; later developments used optical relays[7] and demonstrated entanglement swapping for continuous variables[8]. The teleportation of a quantum state between two single material particles (trapped ions) has now also been achieved[9,10]. Here we demonstrate teleportation between objects of a different nature—light and matter, which respectively represent 'flying' and 'stationary' media. A quantum state encoded in a light pulse is teleported onto a macroscopic object (an atomic ensemble containing $10^{12}$ caesium atoms). Deterministic teleportation is achieved for sets of coherent states with mean photon number ($n$) up to a few hundred. The fidelities are 0.58±0.02 for $n$=20 and 0.60±0.02 for $n$=5— higher than any classical state transfer can possibly achieve[11]. Besides being of fundamental interest, teleportation using a macroscopic atomic ensemble is relevant for the practical implementation of a quantum repeater[2]. An important factor for the implementation of quantum networks is the teleportation distance between transmitter and receiver; this is 0.5 metres in the present experiment. As our experiment uses propagating light to achieve the entanglement of light and atoms required for teleportation, the present approach should be scalable to longer distances.**

Quantum teleportation—a disembodied transfer of a quantum state with the help of distributed entanglement—was proposed in a seminal paper[1]. The generic protocol of quantum teleportation begins with the creation of a pair of entangled objects which are shared by two parties, Alice and Bob. This step establishes a quantum link between them. Alice receives an object to be teleported and performs a joint measurement on this object and her entangled object (a Bell measurement). The result of this measurement is communicated via a classical communication channel to Bob, who uses it to perform local operations on his entangled object, thus completing the process of teleportation.

In our experiment, a pair of entangled objects is created by sending a strong 'in' pulse of light (shown on the left in Fig. 1) through an atomic sample at Bob's location. As a result of the interaction between the light and the atoms the transmitted 'out' light received by Alice's and Bob's atoms become entangled. On Alice's site the entangled pulse is mixed with the pulse to be teleported on a 50/50 beamsplitter (BS in Fig. 1). A Bell measurement in the form of homodyne measurements of the optical fields in the two output ports of the BS is carried out and the results are transferred to Bob as classical photocurrents. Bob performs spin rotations on the atoms to complete the teleportation protocol. Finally, the state of the atoms is analysed to confirm that the teleportation has been successful.

The experiment follows a recent proposal for light-to-atoms teleportation[12] using multimode entanglement of light with an atomic ensemble placed in a magnetic field. We describe teleportation in the language of dimensionless canonical variables[13]; this provides a common description for light and atoms, and allows for a complete tomographic characterization of the states.

The atomic object is a spin-polarized gas sample of approximately $N_{at} = 10^{12}$ caesium atoms in a $25 \times 25 \times 25$ mm paraffin-coated glass cell at around room temperature[14–18] placed in a homogeneous magnetic field (**B**). Atoms are initially prepared in a coherent spin state by a 4-ms circularly polarized optical pumping pulse propagating along the direction of the magnetic field, into the sublevel $F = 4, m_F = 4$ (Fig. 1) of the ground state with the collective ensemble angular momentum $\langle \hat{J}_x \rangle = J_x = 4N_{atoms}$, and the transverse projections with minimal quantum uncertainties, $\langle \delta J_y^2 \rangle = \langle \delta J_z^2 \rangle = \frac{1}{2} J_x$. Changing to the frame rotating at the Larmor frequency $\Omega$ and introducing the canonical variables for the collective transverse atomic spin components[12], we obtain $\hat{X}_A = \hat{J}_y^{rot}/\sqrt{J_x}, \hat{P}_A = \hat{J}_z^{rot}/\sqrt{J_x}$ which obey the canonical commutation relation $[\hat{X}_A, \hat{P}_A] = i$ provided that $J_x \gg \sqrt{\langle \delta J_{y,z}^2 \rangle}, \langle J_{y,z} \rangle$. Here $\hat{X}_A$ and $\hat{P}_A$ are the recipient operators in the teleportation protocol.

The light to be teleported, and the 'in' and 'out' modes (Fig. 1), are described by single mode canonical operators[6,12] $\hat{Y}, \hat{Q}$, and $\hat{y}^{in}, \hat{q}^{in}$ and $\hat{y}^{out}, \hat{q}^{out}$, respectively. These operators obeying $[\hat{Y}, \hat{Q}] = [\hat{y}, \hat{q}] = i$ are quantum analogues of the amplitude and phase of light in classical physics, or, more precisely, of the classical quadrature phase amplitudes $y, q$ in the decomposition of the electric field of light with the frequency $\omega$ as $E \propto y \cos \omega t + q \sin \omega t$ (see Methods for exact definitions). Two non-commuting variables in quantum mechanics cannot be measured without distortion. The challenge of teleportation thus consists of a faithful transfer of these not simultaneously measurable operators, $\hat{Y}, \hat{Q}$, onto atomic operators $\hat{X}_A$ and $\hat{P}_A$. The Raman-type interaction (see Fig. 1 inset) couples the quantum $\omega + \Omega$ sideband of the 'in' field to the Zeeman sublevels separated by the frequency $\Omega = 322\,\text{kHz}$. Therefore we introduce the $\cos \Omega t$, $\sin \Omega t$ components of the light operators $\hat{Y}_{c,s}, \hat{Q}_{c,s}$ and $\hat{y}_{c,s}, \hat{q}_{c,s}$ (see Methods). Canonical operators for the upper sideband mode $\hat{Y}, \hat{Q}$ can be expressed[12] via measurable $\sin(\Omega t)$ and $\cos(\Omega t)$ components, $\hat{Y}_s, \hat{Q}_s, \hat{Y}_c, \hat{Q}_c$, as $\hat{Y} = \frac{1}{\sqrt{2}}(\hat{Y}_s + \hat{Q}_c), \hat{Q} = -\frac{1}{\sqrt{2}}(\hat{Y}_c - \hat{Q}_s)$.

We first describe generation of entanglement between light and atoms. The 'in' strong pulse is $y$-polarized, hence its $x$-polarized mode $\hat{y}^{in}, \hat{q}^{in}$ is in a vacuum state. After interaction with atoms[12], the $x$-polarized 'out' mode operators $\hat{y}^{out}, \hat{q}^{out}$ are given by:

$$\hat{y}_c^{out} = \left\{ \hat{y}_c^{in} + \frac{\kappa^2}{4} \hat{q}_s^{in} + \frac{\kappa^2}{4\sqrt{3}} v_s \right\} + \frac{\kappa}{\sqrt{2}} \hat{P}_A^{in}, \qquad \hat{q}_{s,c}^{out} = \hat{q}_{s,c}^{in}$$

$$\hat{y}_s^{out} = \left\{ \hat{y}_s^{in} - \frac{\kappa^2}{4} \hat{q}_c^{in} - \frac{\kappa^2}{4\sqrt{3}} v_c \right\} - \frac{\kappa}{\sqrt{2}} \hat{X}_A^{in}$$

(1)

The terms in curly brackets in the equations for $\hat{y}$ represent vacuum contributions coming from different orthogonal modes of the 'in' pulse where the canonical operators $v_{s,c}$ represent vacuum temporal higher order canonical modes[12]. The terms containing $\hat{P}_A^{in}$ and $\hat{X}_A^{in}$ describe the imprint of the atomic state on the light via coherent forward scattering from the atomic ensemble, or, in other words, polarization rotation due to the Faraday effect[14,15]. The atomic spin operators are transformed by the interaction with light as follows[12]:

$$\hat{X}_A^{out} = \hat{X}_A^{in} + \frac{\kappa}{\sqrt{2}} \hat{q}_c^{in}, \quad \hat{P}_A^{out} = \hat{P}_A^{in} + \frac{\kappa}{\sqrt{2}} \hat{q}_s^{in} \qquad (2)$$

The second terms in equation (2) describe the imprint of the light state onto atoms via the dynamic Stark effect[14].

The atoms–light entanglement described by equations (1) and (2) is very close[12], under our experimental conditions, to the Einstein–Podolsky–Rosen entanglement optimal for quantum teleportation. The light–atoms coupling constant $\kappa = a_1 \sqrt{N_{ph} N_{at} F \sigma \Gamma / A \Delta} \propto \alpha_0$ has been discussed in detail previously[12,14–18]. (Here $\sigma$ is the dipole cross-section[18], $a_1$ is the vector polarizability[18], $\Gamma = 2.6 MHz$ is the natural linewidth (HWHM) of the transition, $N_{ph} = 4 \times 10^{13}$ is the number of the y-polarized photons in the strong pulse, $\Delta = 825 MHz$ is the blue detuning of light from the atomic resonance, and $A = 4.8 cm^2$ is the cross-section of the atomic sample. As in our previous experiments with the atoms–light quantum interface, strong coupling with the atomic ensemble is achieved in the region of a high resonant optical depth $\alpha_0$. In the experiment we choose a nearly optimal value[12] of $\kappa \approx 1$ by changing $\alpha_0 \propto N_{at}$ with the temperature of the vapour. Note that another condition for strong coherent coupling is a very high $N_{ph}$ in the y-polarized mode.

At Alice's location (Fig. 1), the 'out' pulse is mixed on BS with the object of teleportation—a few-photon x-polarized coherent pulse with frequency $\omega + \Omega$ generated by an electro-optical modulator (EOM). A Bell measurement of canonical variables[6,12,13] is performed by two sets of polarization homodyne detectors in the two output ports of BS (Fig. 1). Homodyne detection followed by the normalization to the vacuum (shot) noise of light[6] is a standard method for measuring canonical variables of light. In our experiment, the strong y-polarized pulse, besides driving the entangling interaction, also plays the role of a local oscillator for the homodyne detection. The variables in phase with the strong pulse $\hat{y}_{c,s} = \frac{1}{\sqrt{2}}(\hat{y}_{c,s}^{out} + \hat{Y}_{c,s})$ are measured via a measurement of the Stokes parameter $\hat{S}_2$ in one output of BS, whereas the out-of-phase components $\hat{q}_{c,s} = \frac{1}{\sqrt{2}}(\hat{q}_{c,s}^{out} - \hat{Q}_{c,s})$ are measured via the Stokes parameter $\hat{S}_3$ in the other arm (see Methods). The $\sin(\Omega t)$ and $\cos(\Omega t)$ components are measured by processing photocurrents with lock-in amplifiers. The Bell measurement of operators $\hat{y}_{c,s}$ and $\hat{q}_{c,s}$ yields four results, $y_{c,s}$ and $q_{c,s}$. Operationally, these values are properly normalized integrals of corresponding photocurrents over the pulse duration (see Methods). As shown in Fig. 1, the photocurrents are combined to yield two feedback signals proportional to $y_s - q_c$ and $y_c + q_s$ which are sent from Alice to Bob. Auxiliary magnetic field pulses[14,17] with frequency $\Omega$ and amplitudes proportional to the feedback signals are applied to the atoms, so that the collective atomic spin variables at Bob's site are shifted to become:

$$\hat{X}_A^{tele} = \hat{X}_A^{out} + g_X(y_s - q_c) = \hat{X}_A^{out} + \frac{1}{\sqrt{2}} g_X(\hat{y}_s^{out} - \hat{q}_c^{out}) + g_X \hat{Y}$$
$$\hat{P}_A^{tele} = \hat{P}_A^{out} - g_P(y_c + q_s) = \hat{P}_A^{out} - \frac{1}{\sqrt{2}} g_P(\hat{y}_c^{out} + \hat{q}_s^{out}) + g_P \hat{Q}$$
(3)

where $g_{X,P}$ are the feedback gains. This step completes the teleportation protocol, as the light operators $\hat{Y}, \hat{Q}$ are now transferred onto atomic operators $\hat{X}_A^{tele}, \hat{P}_A^{tele}$, and all other terms in equation (3) can be made small with a suitable choice of $\kappa$ and $g$.

To prove that we have performed the quantum teleportation, we determine the fidelity of the teleportation. Towards this end, we send a second—verifying—strong pulse of *y*-polarized light through the atomic ensemble after the teleportation is completed. From this measurement we reconstruct the atomic operators $\hat{X}_A^{tele}$ and $\hat{P}_A^{tele}$. The fidelity is the overlap of the input state and the teleported state averaged over the input state distribution[12,14,17]. The classical benchmark fidelity which has to be exceeded in order to claim the success of quantum teleportation is known[11] for a gaussian distribution of coherent states with the width corresponding to the mean photon number $\langle n \rangle$ centred at zero. The experimental fidelity for such distribution can be found as[6,18]:

$$F_n = \frac{2}{\sqrt{(2\langle n \rangle(1-g_X)^2 + 1 + 2\sigma_X^2)(2\langle n \rangle(1-g_P)^2 + 1 + 2\sigma_P^2)}}$$

The gains are defined from the mean values of atomic and light operators: $\bar{X}_A^{tele} = g_X \bar{Y}, \bar{P}_A^{tele} = g_P \bar{Q}$. $\sigma_X^2, \sigma_P^2$ are the variances for the final gaussian state of the atoms.

The mean values for the input light operators are determined from the results of the Bell measurement: $\bar{y}_s - \bar{q}_c = \bar{Y}$ and $\bar{y}_c + \bar{q}_s = \bar{Q}$. The mean values and the variances of the atomic operators are determined from the verifying pulse measurements. Using equations (1) and (3) and the input–output beamsplitter relations[12], we can link the measurement of the verifying pulse on the $S_2$ detector to the atomic mean values: $\bar{y}_c^{ver} = \frac{\kappa}{2} \bar{P}_A^{tele} = \frac{g_P \kappa}{2} \bar{Q}$, $\bar{y}_s^{ver} = \frac{\kappa}{2} \bar{X}_A^{tele} = \frac{g_X \kappa}{2} \bar{Y}$. Using these expressions, we can calibrate $g_{X,P}$, as shown in Fig. 2a where $\hat{y}_c^{ver}$ is plotted as a function of $\hat{Q}$, as the value of $\kappa = 0.93$ is determined independently from the projection noise measurement (see Methods). From the linear fit to this distribution we find $g_P$, which can then be tuned to a desired value electronically. Results plotted in Fig. 2a along with similar results for the other operator $\bar{y}_s(\bar{Y})$ present the proof of the successful classical transfer of the mean values of the quantum mechanical operators $\hat{Y}, \hat{Q}$ of light onto atomic operators.

To verify the success of the quantum teleportation, we have to determine the variances of the two atomic operators which now contain the teleported input light operators. Figure 2b shows an example of results $\hat{y}_c^{ver}, \hat{y}_s^{ver}$ for 250 teleportation runs for a fixed input state. Making use of equation (1) and the beamsplitter relations, we can directly find the atomic state variances from $\text{Var}\{\hat{y}_{s(c)}\}$ of such distribution as $\sigma_{X(P)}^2 = \frac{4}{\kappa^2}\left[\text{Var}\{\hat{y}_{s(c)}\} - \frac{\kappa^4}{48} - \frac{1}{2}\right]$. The final values of $\sigma_X^2, \sigma_P^2$ for a coherent state with a varied phase and a given $\bar{n}$ are found as averages over 10,000 points (that is, 40 runs like in Fig. 2b). For example, for $\bar{n} = 5$ we find $\sigma_{X(P)}^2 = 1.20(1.12)$ taken at gains 0.96 and 0.95 respectively. The results of $\sigma_{X,P}^2(g_{X,P})$ for a range of photon numbers $\bar{n} = 0$ (vacuum), $\bar{n} = 5, 20, 45, 180, 500$ at various gains are summarized in a figure in the Supplementary Methods. From this we obtain $\sigma^2_{X,P}(g_{X,P})$, which can be inserted into the fidelity expression. For a given width of the gaussian distribution of coherent states we find the values of $g_X$, $g_P$, and the corresponding $\sigma^2_{X,P}(g_{X,P})$ which maximize the fidelity. We obtain the following fidelities for distributions with a width $\langle n \rangle = 2, 5, 10, 20, 200$: $F_2 = 0.64 \pm 0.02$; $F_5 = 0.60 \pm 0.02$; $F_{10} = 0.59 \pm 0.02$; $F_{20} = 0.58 \pm 0.02$; $F_{200} = 0.56 \pm 0.03$. The expression for the classical benchmark fidelity[11] $F_n^{class} = \frac{\langle n \rangle + 1}{2\langle n \rangle + 1}$ gives $F_2^{class} = 0.60$; $F_5^{class} = 0.545$; $F_{10}^{class} = 0.52$; $F_{20}^{class} = 0.51$; $F_{200}^{class} = 0.50$ (see Supplementary Methods for details on the fidelity calculations). The maximal $\langle n \rangle$ for successful teleportation is limited by small fluctuations of the classical gain, which for large $\bar{n}$ lead to large uncontrolled displacements of the teleported state with respect to the input state, and hence to the decrease in the fidelity.

In Fig. 3 we show the tomographically reconstructed teleported state with the mean photon number $\bar{n} = 5$. Owing to the gaussian character of the state, the knowledge of the means and the variances of two quadrature phase operators is sufficient for the reconstruction.

Note that the atomic object onto which the teleportation is performed contains hundreds of billions of atoms. However, the number of excitations in the ensemble, of course, corresponds to the number of photons in the initial state of light. Those excitations are coherently distributed over the entire ensemble.

Having demonstrated the teleportation for gaussian states, we now address the applicability of this teleportation protocol to the teleportation of a light qubit, which is relevant for, for example, quantum computing[3]. In the Supplementary Notes we give the derivation of the predicted qubit fidelity, $F_q$, based on the performance of our teleportation protocol for coherent states. For experimentally relevant values of losses and decoherence, $F_q = 0.72$—higher than the best classical fidelity for a qubit of $0.67$— can be predicted. In order to experimentally demonstrate such qubit teleportation, a source generating such a qubit in a temporal, spectral and spatial mode compatible with our atomic target is required. First steps towards generation of an atom-compatible qubit state of light have been recently made using atomic ensembles[19–21], single atoms in a cavity[22,23], and a photon subtracted squeezed state[24].

In our experiment, the entanglement generation and the Bell measurement overlap in time because the duration of the strong pulse and the pulse to be teleported is 2 ms, which is much longer than the time it takes light to travel from Alice to Bob. This situation, also the case in some teleportation experiments[6,8], is different, for example, from the teleportation[7,9,10] in which the entanglement generation and the Bell measurement are separated in time. This feature is not inherent to our teleportation scheme—indeed, in principle, a shorter strong pulse (of higher power) would generate the same entanglement on a timescale short compared to the propagation time, especially if the distance from Alice to Bob is extended to a few kilometres. The teleportation distance can be increased, and is limited only by propagation losses of light and the atomic coherence lifetime. The timing of the entanglement generation and the Bell measurement may be potentially important for future applications.

Further improvement of the present teleportation protocol can be achieved by performing more complex photocurrent processing with the same homodyne set-up. As shown in ref. 12 and in the Supplementary Notes, a fidelity of 0.93 can be achieved if such processing is combined with the use of an experimentally feasible[25] 6 dB squeezed strong pulse.

## Methods

**Calibration and measurement techniques.**

Physically, we perform measurements of the Stokes operators of light by two sets of balanced homodyne detectors (Fig. 1). The measurements on the first pulse represent the generalized Bell measurement. The same measurements on the second (verifying) pulse allow us to determine the teleported atomic state by performing quantum state tomography. The relevant $\cos(\Omega t)$ and $\sin(\Omega t)$ modulation components of the Stokes operators are measured by processing the corresponding photocurrents with lock-in amplifiers. The Stokes operators of interest are $\hat{S}_2$ (which is the difference between photon fluxes in the modes polarized at $\pm 45°$ to the vertical axis, and $\hat{S}_3$ (which is the corresponding quantity for the left- and right-hand circular polarizations).

Calibration of the measurement of canonical variables for light is based on measurements of the shot (vacuum) noise level. We measure the Stokes parameters for the *x*-polarization mode in a vacuum state. The linear dependence of the variance of the measured photocurrents on the optical power of the strong pulse proves that the polarization state of light is, in fact, shot (vacuum) noise limited[25]. All other measurements of $\hat{S}_2, \hat{S}_3$ are then normalized to this shot noise level, yielding the canonical variables as

$$y_c = \frac{1}{\sqrt{2}\int_0^T d\tau \cos(\Omega t) S_2^{vacuum}(\tau)} \int_0^T d\tau \cos(\Omega t) S_2(\tau)$$

and similarly for $q_c(S_3)$ and the $\sin(\Omega t)$ components. Since our detectors have nearly unity (better than 0.97) quantum efficiency, the Stokes operators can be operationally substituted with measured photocurrents.

Next we need to calibrate the atomic coherent (projection) noise level. Whereas balanced homodyne detection for light has become an established technique for determination of the vacuum state[6], a comparable technique for atoms is a relatively recent invention. Here we utilize the same procedure as used in our previous experiments on the atoms–light quantum interface[14,15]. We use the fact that the vacuum (projection) noise level for collective atomic spin states in the presence of a bias magnetic field can be determined by sending a pulse of light through two identical atomic ensembles with oppositely oriented macroscopic spin orientation. We therefore insert a second atomic cell in the beam. As described in detail in ref. 15, the transmitted light state in this experiment is given by

$$\hat{y}_c^{out} = \hat{y}_c^{in} + \frac{\kappa}{\sqrt{2}}\left(\hat{P}_{atom1} + \hat{P}_{atom2}\right) = \hat{y}_c^{in} + \kappa \hat{P}_{total}$$

where $\hat{P}_{total}$ is the spin canonical variable for the entire 2-cell atomic sample. Intuitively this equation can be understood by noting that terms proportional to $\kappa^2$ in equation (1) cancel out for propagation through two oppositely oriented ensembles. A similar equation holds for $\hat{y}_s^{out}$ with substitution of $\hat{X}_{total}$ for $\hat{P}_{total}$. The results for $\text{Var}\left(\hat{y}_{c,s}^{out}\right)$ as a function of the number of atoms are shown in the figure in the Supplementary

Methods. The fact that the points lie on a straight line, along with the independent measurement of the degree of spin polarization above 0.99, proves[14,15,18] that we are indeed measuring the vacuum (projection) noise of the atomic ensemble. $\kappa^2$ for different atomic numbers is then calculated from the graph (Supplementary Methods). Its values are in good agreement with the theoretical calculation[18] according to $\kappa = a_1\sqrt{N_{ph}N_{at}F\sigma\Gamma/A\Delta}$. In the experiment, we monitor the number of atoms by sending a weak off-resonant probe pulse along the direction $x$ and measuring the Faraday rotation angle proportional to the collective macroscopic spin of the ensemble $J_x = 4N_{atoms}$. This Faraday angle is monitored throughout the teleportation experiment, so that the value of $\kappa^2$ is known at every stage.

**Decoherence and losses.**

The main sources of imperfections are decoherence of the atomic state and reflection losses of light. For experimental values of the atomic decoherence and losses, the model developed in ref. 12 predicts, for example, $F_5 = 0.66$, which is still higher than the observed value owing to imperfections unaccounted for in the model but comparable to the obtained experimental results. Dissipation also affects the experimental state reconstruction procedure. The main effect of the light losses $\varepsilon = 0.09$ is that it modifies $\kappa$ into $\kappa\sqrt{1-\varepsilon}$. However, this modified $\kappa$ is, in fact, exactly the parameter measured in the two-cell calibration experiment described above, so no extra correction is due because of these losses. There is also a small amount of electronic noise of detectors which can be treated as an extra vacuum contribution to the input state.

**Standard deviation of the teleportation fidelity.**

The standard deviation of the fidelity for $\langle n \rangle \leq 20$ is calculated as follows:

$$SD(F) = \sqrt{\delta_{PN}^2 + \delta_{SN}^2 + \delta_{el}^2 + \delta_{\beta}^2 + \delta_{SNR}^2 + \delta_{fit}^2 + \delta_{g}^2} =$$
$$= 10^{-2}\sqrt{1.0^2 + 1.65^2 + 0.1^2 + 0.3^2 + 0.2^2 + 1.2^2 + 0.8^2} \approx 0.02$$

where $\delta_{PN} = 0.01$ is the contribution to the $SD(F)$ due to the projection noise fluctuations including an error due to imperfect optical pumping, $\delta_{SN} = 0.017$ is the contribution due to the shot noise level uncertainty, $\delta_{el} = 0.001$ is the contribution of the electronics noise level fluctuations, $\delta_\beta = 0.003$ is the uncertainty due to fluctuations in the atomic decay constant, $\delta_{SNR} = 0.002$ is the contribution of the fluctuations in the ratio of responses of two pairs of detectors, $\delta_{fit} = 0.012$ is the deviation due to the uncertainty of the quadratic fit of the atomic noise as a function of gain, and $\delta_g = 0.008$ is the contribution of the gain fluctuations. For $\langle n \rangle > 20$, $\delta_{fit} = 0.016$, giving $SD(F) \approx 0.026 \approx 0.03$.

1.  Bennett, C. H. *et al.* Teleporting an unknown quantum state via dual classical and Einstein-Podolsky-Rosen channels. *Phys. Rev. Lett.* **70,** 1895–1899 (1993).

2.  Briegel, H. J., Dur, W., Cirac, J. I. & Zoller, P. Quantum repeaters: the role of imperfect local operations in quantum communication. *Phys. Rev. Lett.* **81,** 5932–5935 (1998).

3.  Gottesman, D. & Chuang, I. Demonstrating the viability of universal quantum computation using teleportation and single-qubit operations. *Nature* **402,** 390–393 (1999).

4.  Bouwmeester, D. *et al.* Experimental quantum teleportation. *Nature* **390,** 575–579 (1997).

5.  Boschi, D., Branca, S., De Martini, F., Hardy, L. & Popescu, S. Experimental realization of teleporting an unknown pure quantum state via dual classical and Einstein-Podolsky-Rosen channels. *Phys. Rev. Lett.* **80,** 1121–1125 (1998).

6.  Furusawa, A. *et al.* Unconditional quantum teleportation. *Science* **282,** 706–709 (1998).

7.  de Riedmatten, H. *et al.* Long distance quantum teleportation in a quantum relay configuration. *Phys. Rev. Lett.* **92,** 047904 (2004).

8.  Takei, N., Yonezawa, H., Aoki, T. & Furusawa, A. High-fidelity teleportation beyond the no-cloning limit and entanglement swapping for continuous variables. *Phys. Rev. Lett.* **94,** 220502 (2005).

9.  Barrett, M. D. *et al.* Deterministic quantum teleportation of atomic qubits. *Nature* **429,** 737–739 (2004).

10. Riebe, M. *et al.* Deterministic quantum teleportation with atoms. *Nature* **429,** 734–737 (2004).

11. Hammerer, K., Wolf, M. M., Polzik, E. S. & Cirac, J. I. Quantum benchmark for storage and transmission of coherent states. *Phys. Rev. Lett.* **94,** 150503 (2005).

12. Hammerer, K., Polzik, E. S. & Cirac, J. I. Teleportation and spin squeezing utilizing multimode entanglement of light with atoms. *Phys. Rev. A* **72,** 052313 (2005).

13. Vaidman, L. Teleportation of quantum states. *Phys. Rev. A* **49,** 1473–1476 (1994).

14. Julsgaard, B., Sherson, J., Fiurášek, J., Cirac, J. I. & Polzik, E. S. Experimental demonstration of quantum memory for light. *Nature* **432,** 482–486 (2004).

15. Julsgaard, B., Kozhekin, A. & Polzik, E. S. Experimental long-lived entanglement of two macroscopic objects. *Nature* **413,** 400–403 (2001).

16. Julsgaard, B., Schori, C., Sørensen, J. L. & Polzik, E. S. Atomic spins as a storage medium for quantum fluctuations of light. *Quant. Inf. Comput.* **3**(special issue)**,** 518-534 (2003).

17. Julsgaaard, B., Sherson, J., Sørensen, J. L. & Polzik, E. S. Characterizing the spin state of an atomic ensemble using the magneto-optical resonance method. *J. Opt. B* 6, 1, 5–14 (2004).

18. Sherson, J., Julsgaaard, B. & Polzik, E. S. Deterministic atom-light quantum interface. *Adv. At. Mol. Opt. Phys.* (in the press); preprint at <http://arxiv.org/quant-ph/0601186> (2006).

19. Chou, C. W., Polyakov, S. V., Kuzmich, A. & Kimble, H. J. Single-photon generation from stored excitation in an atomic ensemble. *Phys. Rev. Lett.* **92,** 213601 (2004).

20. Chaneliere, T. *et al.* Storage and retrieval of single photons transmitted between remote quantum memories. *Nature* **438,** 833–836 (2005).

21. Eisaman, M. D. *et al.* Electromagnetically induced transparency with tunable single-photon pulses. *Nature* **438,** 837–841 (2005).

22. Kuhn, A., Hennrich, M. & Rempe, G. Deterministic single-photon source for distributed quantum networking. *Phys. Rev. Lett.* **89,** 067901 (2002).

23. McKeever, J. *et al.* Deterministic generation of single photons from one atom trapped in a cavity. *Science* **303,** 1992–1994 (2004).

24. Neergaard-Nielsen, J. S., Melholt Nielsen, B., Hettich, C., Mølmer, K. & Polzik, E. S. Generation of a superposition of odd photon number states for quantum information networks. *Phys. Rev. Lett.* (in the press); preprint at <http://arxiv.org/quant-ph/0602198> (2006).

25. Polzik, E. S., Carri, J. & Kimble, H. J. Spectroscopy with squeezed light. *Phys. Rev. Lett.* **68,** 3020–3023 (1992).

**Acknowledgements** The experiment was performed at the Niels Bohr Institute, and was funded by the Danish National Research Foundation through the Center for Quantum Optics (QUANTOP), by EU grants COVAQIAL and QAP, and by the Carlsberg Foundation. I.C. and E.S.P. acknowledge the hospitality of the Institut de Ciències Fotòniques (ICFO) in Barcelona where ideas leading to this work were first discussed. The permanent address of K.H. is the Institut für theoretische Physik, Innsbruck, Austria.

**Author Information** Reprints and permissions information is available at www.nature.com/reprints. The authors declare no competing financial interests. Correspondence and requests for materials should be addressed to E.S.P. (polzik@nbi.dk).

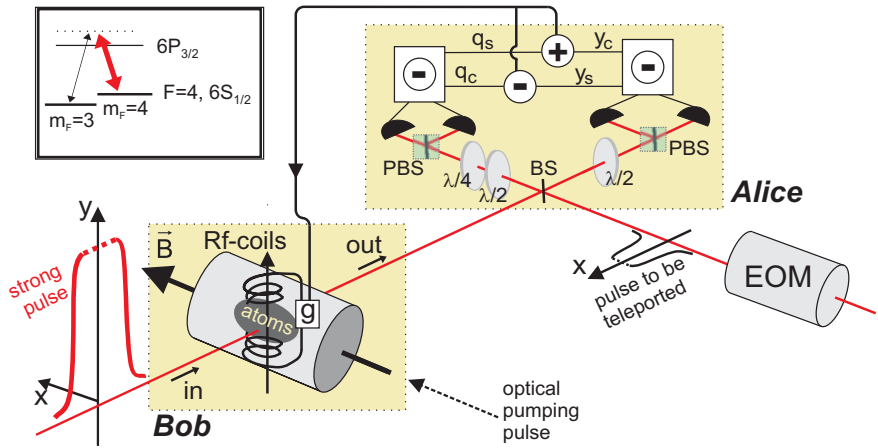

**Figure 1 Experimental set-up for teleportation of light onto an atomic ensemble.**
Atoms are initially optically pumped into $F=4$, $m_F=4$ state with a 4-ms pulse. A strong
$y$-polarized 2-ms 'in' pulse of light is then sent through the atomic sample at Bob's
location and becomes entangled with the atoms (the pulse length is around 600 km and
is not shown to scale in the figure). The pulse travels 0.5 m to Alice's location, where it
is mixed on a beamsplitter (BS) with the object of teleportation—a few-photon coherent
pulse of light—generated by the electro-optical modulator (EOM) synchronously with
the strong pulse. In the two output ports of the BS, two polarization beamsplitters (PBS)
split light onto two pairs of detectors which perform a polarization homodyne
measurement (a Bell measurement). The results of these measurements are combined,
processed electronically, as described in the text, and sent via a classical communication
channel to Bob. There they are used to complete the teleportation onto atoms by shifting
the atomic collective spin state with a pulse of a radio-frequency (RF) magnetic field of
0.2-ms duration. After a delay of 0.1 ms, a second strong pulse—the verifying pulse—is
sent to read out the atomic state, in order to prove the successful teleportation. Inset,
relevant atomic sublevels and light modes (not to scale). The frequency difference
between a weak quantum field (black arrow) and the strong entangling field (thick red
arrow) is equal to the Zeeman splitting of the ground state sublevels.

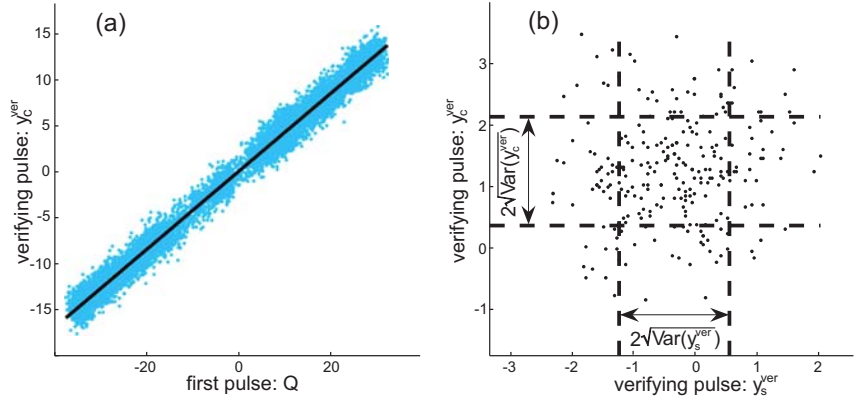

**Figure 2 Raw experimental data for a series of teleportation runs. a**, Calibration of the teleportation feedback gain. Verifying pulse canonical variable $y_c^{ver}$ versus the input pulse canonical variable $Q$ for 10,000 teleportation runs. All dimensionless canonical variables are normalized so that their variance for a vacuum state is 1/2. The coherent input state used in the plot has a mean photon number of $\bar{n} \approx 500$, and is slowly modulated in phase during this measurement. The straight line fit is used for calibration of the feedback gain (see comments in the text). **b**, An example of data from which the atomic state variances after the teleportation are determined. Two canonical variables of the verifying pulse, $y_c^{ver}$ and $y_s^{ver}$, are plotted for an input state with $\bar{n} = 5$ and a fixed phase. The dashed lines indicate twice the standard deviation intervals $2\sqrt{\mathrm{Var}\left(y_{c,s}^{ver}\right)}$ which are used to determine the atomic state variances as discussed in the text.

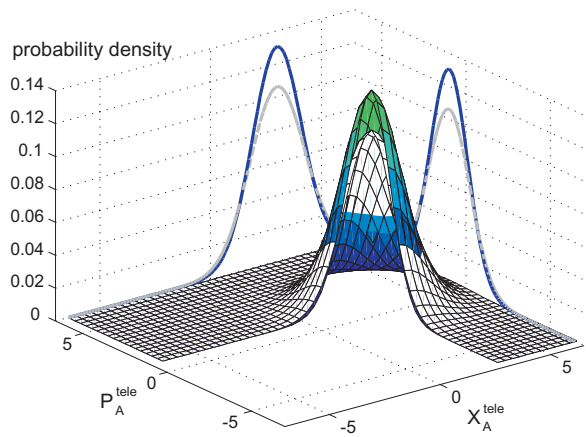

**Figure 3 Tomographic reconstruction of a teleported state with $\bar{n} = 5$ (coloured contour) versus the state corresponding to the best classical state transfer.**
Canonical variables plotted on horizontal axes are normalized so that their variance for a vacuum state is 1/2.

**Supplementary Methods.**
**Atomic state variances, optimization of classical gains, and the fidelity calculation**

Fidelity is calculated for a Gaussian distribution of coherent states centered at zero with the width $\langle n \rangle$ because a strict classical bound is known for such distributions. We have performed the experiment described in the paper for various photon numbers $\bar{n} = 0$ (vacuum), and $\bar{n} = 5, 20, 45, 180, 500$. Empirically we found that the reconstructed atomic variances could be grouped into two sets, one with $\bar{n} \leq 20$ and another with $\bar{n} > 20$. Within each set the variance was independent on the input state, but for the set containing higher photon numbers the value of the variance was slightly higher for technical reasons. Thus, when estimating the fidelity of the teleportation for $\langle n \rangle \leq 20$ we use the set of measurements for low photon numbers only, whereas for higher photon numbers we have to include both sets. For each set we repeated the experiment for various values of the gain. The results for $\langle n \rangle \leq 20$ are shown in the figure:

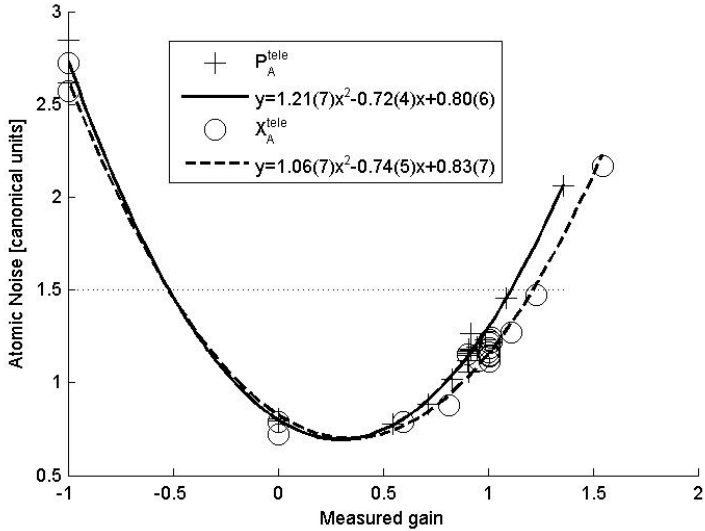

The uncertainties stated in the fit are 95% confidence intervals. Theory predicts a quadratic dependence of the atomic variance on the gain (Ref. 12) which is consistent with the experimental data as shown above. Using the quadratic fit to the data we obtain $\sigma_{X,P}^2(g_{X,P})$ which can be inserted into the expression for the fidelity corresponding to a Gaussian distribution of coherent states with the width $\langle n \rangle$:

$$F = \frac{2}{\sqrt{(2\langle n \rangle(1-g_X)^2 + 1 + 2\sigma_X^2(g_X))(2\langle n \rangle(1-g_P)^2 + 1 + 2\sigma_P^2(g_P))}} \quad .$$

It is of course crucial that the variances are independent of $\langle n \rangle$. The fidelity is now a function of $\langle n \rangle, g_X$, and $g_P$ only, which can easily be optimized with respect to $g_X$ and $g_P$ yielding the optimized experimental fidelity vs. the width of the input state distribution $\langle n \rangle$. In the figure below we show the result of this optimization (full-drawn) together with the classical boundary (dashed):

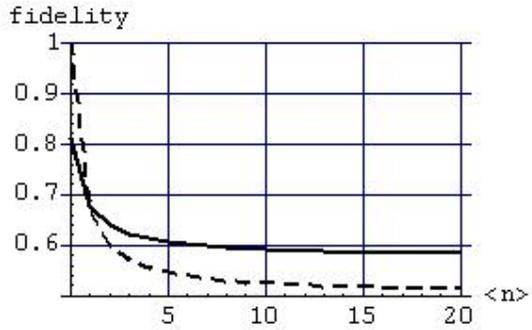

As can be seen the experimental fidelity is higher than the classical for $\langle n \rangle \geq 2$. Below we also show the optimal gain for each $\langle n \rangle$:

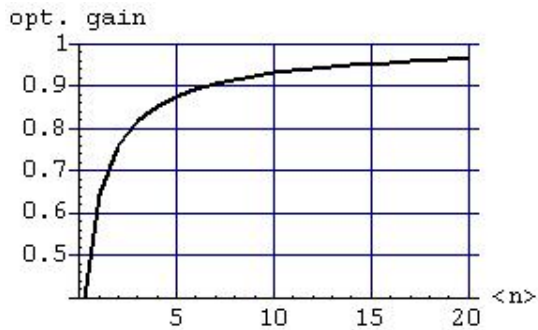

As can be seen from the graph, the optimal gain approaches unity for high photon numbers but is significantly lower than unity for low photon numbers. The reason is that for low photon numbers a more prominent role is played by the vacuum contribution (which is perfectly transferred for zero gain and $\kappa = 0$).

For $\bar{n} > 20$ the atomic variance is larger so for the calculation of the fidelity for $\langle n \rangle > 20$ we conservatively choose to calculate $\sigma^2_{X,P}(g_{X,P})$ from all the data:

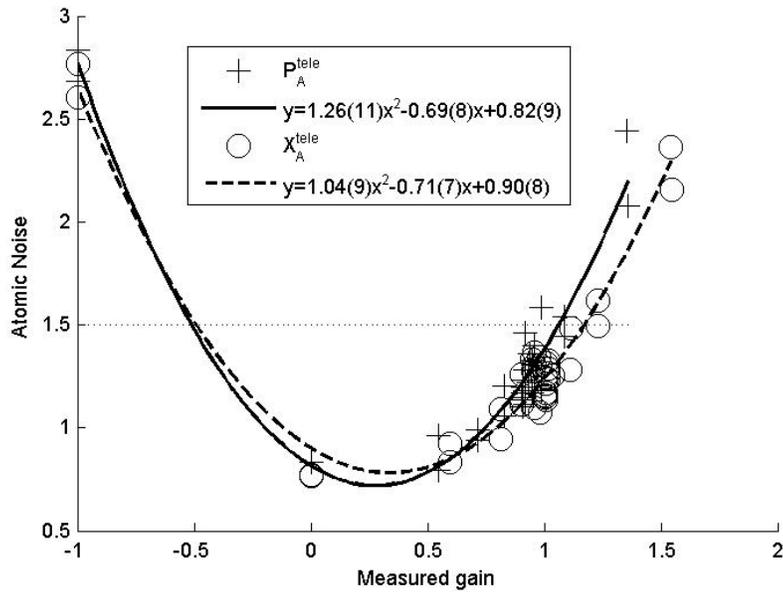

We note that the reconstructed variances are higher and the scattering of points is somewhat larger. In the same way as above the fidelity can be optimized for different photon numbers.

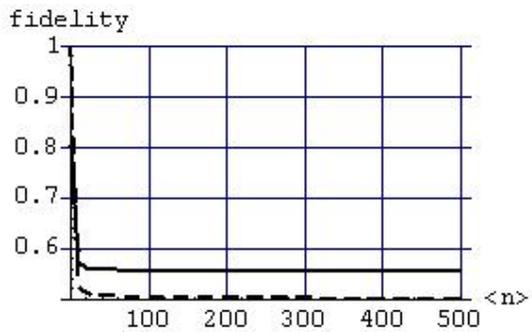

As can be seen the experimental fidelity quickly saturates around 55.5%. Below we show that the optimal gain also approaches unity rather fast.

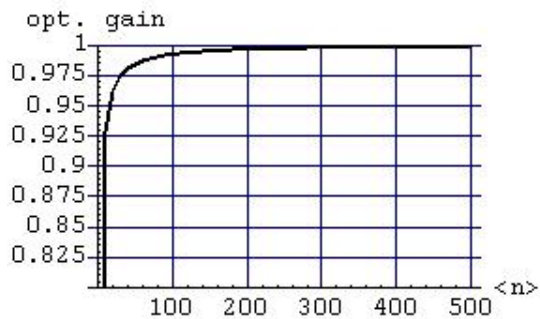

**Projection noise measurement and determination of the coupling constant κ.**

The projection noise (coherent spin state noise) and the coupling constant $\kappa$ are determined in a two cell experiment, as in Ref. 14,15,18. The projection noise is, as described in the paper, the last term in the expression for the state of the transmitted light $\hat{y}_c^{out} = \hat{y}_c^{in} + \frac{\kappa}{\sqrt{2}}\left(\hat{P}_{atom1} + \hat{P}_{atom2}\right) = \hat{y}_c^{in} + \kappa \hat{P}_{total}$ in a two cell experiment. The projection noise value and the coupling constant $\kappa$ can be found as $\kappa^2 = 2\left(Var\{\hat{y}_c^{out}\} - Var\{\hat{y}_c^{in}\}\right) = 2 Var\{\hat{y}_c^{out}\} - 1$, where we took into account that $Var\{\hat{y}_c^{in}\} = Var\{\hat{P}_{total}\} = \frac{1}{2}$. Operationally $Var\{\hat{y}_c^{in}\}$ is the variance of the transmitted probe light normalized to the variance of the shot noise of the probe minus unity. Measuring the noise of the transmitted probe $Var\{\hat{y}_c^{out}\}$ as a function of the number of atoms (more precisely as a function of a Faraday rotation angle of an auxiliary probe proportional to the number of atoms) we can thus find $\kappa^2$ for a given number of atoms. The results of such measurements are shown in the Figure.

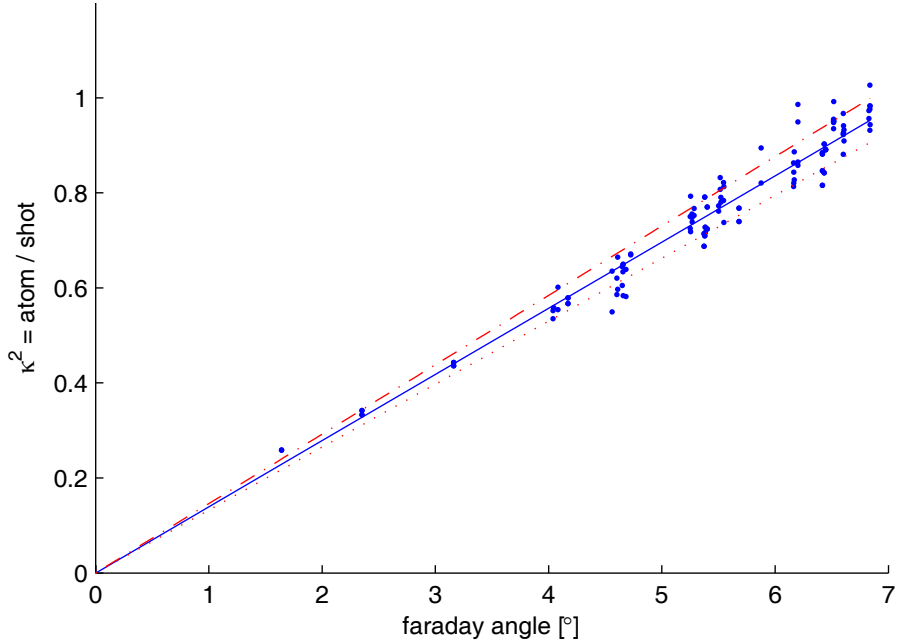

Several experimental runs taken over a time period of a few weeks are collected in the figure to show a good reproducibility of the projection noise level. The fact that the points lie on a straight line is in agreement with our theory since for the projection noise (coherent spin state noise) $Var\{\hat{P}_{total}\} = \frac{1}{2}$ and
$\kappa^2 = 2\left(Var\{\hat{y}_c^{out}\} - Var\{\hat{y}_c^{in}\}\right) \propto N_{Atoms}$

The value of $\kappa^2$ for a given value of the Faraday angle needed for the atomic state reconstruction can be directly read from the figure. The dashed-dotted and dotted lines are the fits used to determine the standard deviation of the projection noise slope.

**Atomic decoherence**

The rate of atomic decoherence in a paraffin coated glass cell in the absence of interaction with light is very low[15,17,18] corresponding to the coherence lifetime of $40 m\sec$. However, the verifying pulse causes a much faster decoherence of the atomic state via the process of light-induced collisions[14,18]. This leads to the reduction in the mean spin values $J_{y,z}$ according to $e^{-\beta\tau}$. We must adjust the gain calibration to take this small but still important effect into account. The decay constant $\beta = 0.09 m\sec^{-1}$ of the mean atomic spin orientation in the presence of the probe light is measured in a separate experimental run. As a result of this decay the verifying pulse measures reduced mean values $\bar{y}_c = \frac{1}{2} g\kappa e^{-\beta\tau}\bar{Q}, \bar{y}_s = \frac{1}{2} g\kappa e^{-\beta\tau}\bar{Y}$ as compared to the teleported mean values, where $\tau = 1 m\sec$ is the time interval from the beginning of the verifying pulse to its center yielding $e^{-\beta\tau} = 0.91$. Thus the unity gain $g$ is determined from the condition that the slope in Fig.2a is
$$\frac{\bar{y}_c}{\bar{Q}} = \frac{1}{2} g\kappa e^{-\beta\tau} = \frac{1}{2} \cdot 1 \cdot 0.93 \cdot 0.91 = 0.42.$$

**Supplementary Notes**

**Calculation of the fidelity for a qubit teleportation and a protocol with improved fidelity**

In this section we show (i) how one can relate the teleportation fidelity of coherent states to that of qubits and (ii) how a fidelity approaching unity can be achieved in a more sophisticated teleportation protocol.

(i) Let us call $E$ the (completely positive) map that transforms the state to be teleported to the teleported one. We assume that we know the action of $E$ on coherent states, i.e., $E(|\alpha\rangle\langle\alpha|)$. The goal is to determine the qubit fidelity, $F_q$, of this map. This is given by

$$F_q = \int d\Omega \; \langle\Psi(\Omega)|E(|\Psi(\Omega)\rangle\langle\Psi(\Omega)|)|\Psi(\Omega)\rangle$$

where $|\Psi(\Omega)\rangle = \cos(\theta/2)|0\rangle + e^{i\varphi}\sin(\theta/2)|1\rangle$ and the integration is over the $4\pi$ solid angle. This expression can be easily determined in terms of

$$\langle a|E(|n\rangle\langle m|)|b\rangle = \frac{1}{\sqrt{n!m!}}\partial_\alpha^n \partial_{\alpha^*}^m \left[e^{|\alpha|^2}\langle a|E(|\alpha\rangle\langle\alpha|)|b\rangle\right]_{\alpha=0}$$

with $n,m,a,b=0,1$. Note that, in practice, one can also determine these quantities in terms of other measurable quantities. In the present experiment, one can characterize $E$ as

$$E(|\alpha\rangle\langle\alpha|) = \frac{1}{2\pi s^2}\int d^2\beta\, e^{-|\beta-g\alpha|^2/(2s^2)}|\beta\rangle\langle\beta|$$

where $s^2 = 4\sigma^2 - 1$ is related to the atomic variance, $\sigma^2 = Var(X_A^{tele}) = Var(P_A^{tele})$ and $g$ is the gain (whose values are restricted given the complete positiveness of $E$). For this map we obtain

$$F_q = \frac{6 + 16s^2 + 24s^4 + 4(g-1)(1-2s^2) + (g-1)^2(1-6s^2)}{6(1+2s^2)^3}$$

This shows that, in principle, one can obtain arbitrarily high fidelities for $g=1$ and small variances. For a particular teleportation protocol used in this paper, the value of $\sigma^2$ follows from the expressions for $X_A^{tele}, P_A^{tele}$ (see text) and one obtains a theoretical limit on the fidelity of 0.74 for $\kappa=1$, in the absence of losses and decoherence. Including 10% of light losses and atomic decay as in Ref. 12 of the paper, the theoretical prediction is still 0.72 for the same value of $\kappa$.

(ii) A fidelity approaching unity can be achieved, in principle, by using squeezed light in the entangling pulse and measuring a number of higher order scattering modes in addition to the zero-th order cosine and sine modes. As explained in Ref .12 of the paper, the input-ouput relations describing the state of atoms and light after the interaction can be appended by similar relations for higher order modes $\left[y_{\alpha,n}, q_{\beta,m}\right] = i\delta_{\alpha\beta}\delta_{nm}$ where $\alpha,\beta = c,s$ and $n,m = 1,2,\ldots$ . Note that for the noise operators $v_\alpha$ in Eq. (1) we have $v_\alpha = q_{\alpha,1}$. The envelopes of these modes are, apart from the (co)sine factor, given by the $n$th-order Legendre Polynomials (see Ref. 12 of the paper). Note that they can be measured by simply multiplying the photocurrents recorded in the polarization homodyne measurements of the present setup by suitable slowly varying time envelopes. In the light-atom interaction, these modes are transformed as $q^{out}_{c(s),n} = q^{in}_{c(s),n}$ and $y^{out}_{c(s),n} = y^{in}_{c(s),n} \pm \frac{\kappa^2}{4}\left(c_n q^{in}_{s(c),n-1} - c_{n+1} q^{in}_{s(c),n+1}\right)$ where $c_n = \left(4n^2-1\right)^{-1/2}$. Extension of the teleportation protocol which includes these modes amounts to preparing the input state in some appropriate linear combination thereof, that is, in a mode $Y = \sum_n g_n \frac{1}{\sqrt{2}}(Y_{s,n} + Q_{c,n})$ and $Q = -\sum_n g_n \frac{1}{\sqrt{2}}(Y_{c,n} - Q_{s,n})$ where $\sum_n g_n^2 = 1$. Optimization with respect to the coefficients $g_n$ shows that it suffices to include the first three modes only ($g_n = 0$ for $n > 3$) in order to achieve a final state of atoms after the feedback which is close to $X_A^{tele} = Y + \frac{1}{\sqrt{2}}\sum_n g_n y^{in}_{s,n}$ and $P_A^{tele} = Q + \frac{1}{\sqrt{2}}\sum_n g_n y^{in}_{c,n}$. The sums in these expressions amount to half a unit of vacuum noise in each spin component or a teleportation fidelity of $F = 0.80$. The corresponding optimal coefficients $g_n$ determine the envelope of the input mode $Y, Q$, which is a slowly decaying profile, as shown in Figure 1.

The remaining half unit of added noise is due to the vacuum fluctuations of the entangling beam. It is possible to reduce this remaining noise by using squeezed light, such that the fidelity would approach unity, as the variances of $y^{in}_{\alpha,n}$ approach zero. The squeezing spectrum should be broad enough to cover the sidebands at $\pm 322 kHz$, which is trivial with existing squeezing sources, for example squeezing of 6 dB was observed within the bandwidth of 2 MHz in Ref. 25 of the paper. Squeezing of 6dB (10dB) would yield the fidelity of $F = 0.93$ ($F = 0.96$). Details about these improved protocols will be presented elsewhere.

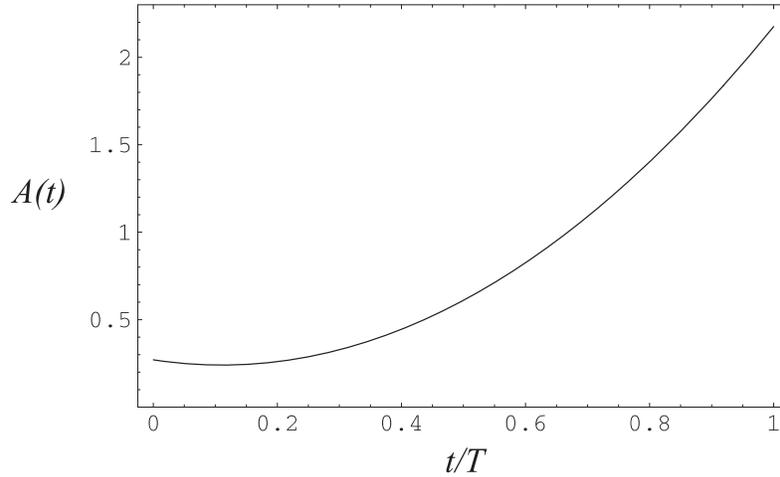

**Figure 1:** Optimal pulse envelope *A(t)* for the input state versus time *t* in units of pulse length *T* for a coupling strength of $\kappa = 2.3$ and light squeezing of 6dB in an improved teleportation protocol including the first three scattering modes. The corresponding fidelity is 0.93. (The square integral of *A(t)* is normalized to 1.)